# An entertaining resonance experiment with just two spring scales

Yajun Wei

Guangzhou University

runnerwei@qq.com

## 1. Introduction

Resonance is a topic included in most introductory physics courses. Any mechanical system experiences resonance if it is driven by a periodic force with a frequency that matches its natural frequency. There are plenty of simple demonstrations of the resonance phenomena of mechanical systems which can be set up using readily available items [1-5]. Here we present a very simple approach to demonstrate the phenomena using just two spring scales. The experiment presented here performs a "frequency sweep" and is also very entertaining to watch.

## 2. The demonstration

In our experiment, two spring scales are hung hook-to-hook to form a mechanical system exhibiting vibration that is easily visible, as shown in Fig. 1. A person holds the top of one spring scale and walks around the classroom varying their pace. The periodic motion of the demonstrator's body while walking forces the Newton meter system to oscillate. When the demonstrator walks at a particular pace, the vibration of the system becomes quite vigorous, as shown in Fig. 1b, while at any pace that is much slower (Fig. 1a) or much faster (Fig. 1c) than this resonance pace, the amplitude of the vibration of the system is apparently smaller. One moves around the classroom at a gradually increasing pace and by doing so actually performs a "frequency sweep".

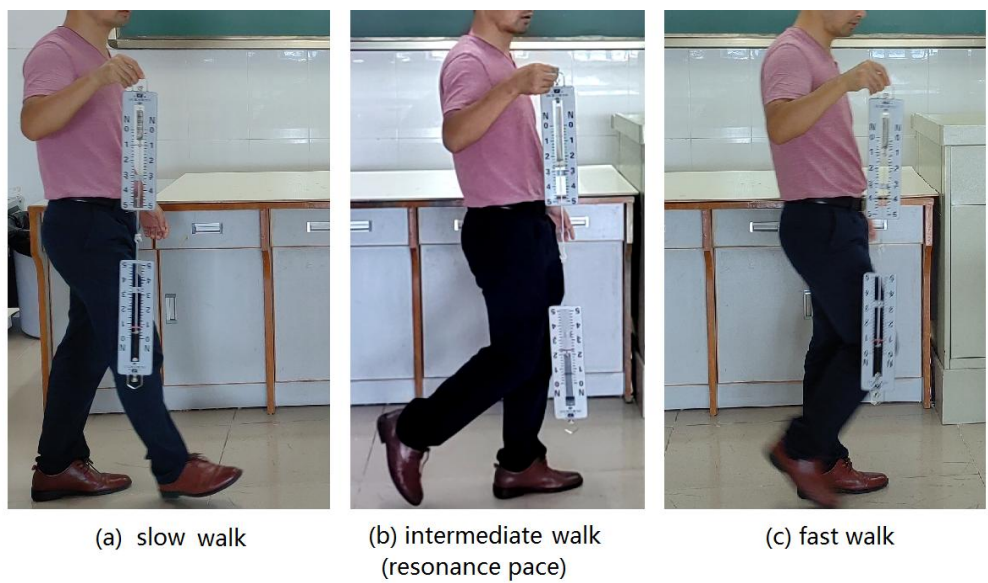

(a) slow walk   (b) intermediate walk (resonance pace)   (c) fast walk

**Fig. 1. Screenshots of a video of the demonstration performed in a classroom.**

In a frequency sweep, one measures a physical quantity at a certain frequency and then varies the frequency to take another measurement. This process is repeated for a range of different frequencies with the frequencies being increased (or decreased) gradually. In science and engineering, "frequency sweep" is a very widely used experimental technique for probing systems and materials with frequency dependent properties [6-10]. Therefore this simple demonstration offers the added benefit of introducing this important concept in an elementary physics course.

I have carried out this demonstration in front of three different classes. On each occasion, when I walked around the classroom at resonance pace with the system vibrating vigorously and when I walked at a quick pace, the students laughed out loudly. Because it is entertaining to watch, the experiment impresses onlookers and this makes it a great demonstration to use as part of a science show.

## 3. Quantitative Experiment

A quantitative experiment can also be performed with this simple system of two spring scales. The walk can be filmed by using a camera/mobile phone and the video analyzed to quantitatively determine the frequency of the driving force and the amplitude of the vibration at different walking paces. Thereafter one can plot the amplitude vs driving frequency curve for the system.

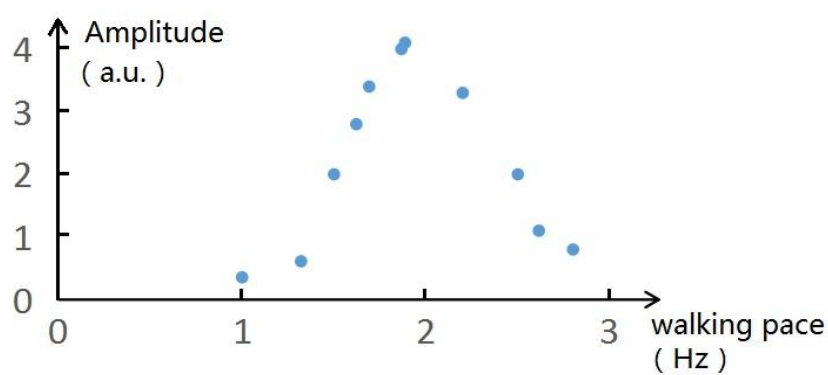

**Fig. 2. Amplitude vs. driving frequency (walking pace) data extracted from a video.**

Fig. 2 shows an amplitude vs driving frequency curve with data extracted from a video. Here the amplitude is measured as the difference between the maximum vertical separation between the two spring scales during oscillation and the separation while at rest. This is taken by using a ruler to measure the distance directly from the computer screen while playing back the video frame by frame to find the maximum. While being plotted, the unit is taken as an arbitrary unit for amplitude so one does not need to worry about the zooming rate.

Walking, even along a straight line, has an oscillatory aspect. The oscillation is caused by the periodic motion of the feet. Therefore we believe that the vertical motion of the feet (on and off the ground) is the driving force for the spring scales to oscillate. We propose the hypothesis that the time taken to move one step should be the period of the oscillation associated with walking (also the period of the driving force). So for the horizontal axis, the driving frequency is taken as $N/\Delta t$ where $N$ is the number of steps during the time interval $\Delta t$. The value of $\Delta t$ is typically a couple of seconds during which the walking speed is maintained as far as possible at a constant pace.

From Fig. 2 one can see that resonance occurs when the driving frequency is about 1.9 Hz, which corresponds to a walking pace of roughly two steps per second. In our experiment, the mass and stiffness constant of the spring scales are measured at 95g and 22.6 N/m respectively. The effective stiffness constant of two such springs connected in series is then 11.3N/m. Therefore we are able to calculate the natural frequency of the two spring scale system to be

$$T = 2\pi\sqrt{\frac{m}{k}} = 0.55s,$$

which corresponds to a frequency of 1.8 Hz. This value is very close to the measured resonance frequency of 1.9 Hz. The result proves our hypothesis about the origin of the driving force and its period.

## 4. The Criteria

It was an accidental discovery that walking with two spring scales could result in an apparent resonance while the spring scales were used for another purpose. Later on, our analysis to trace the underlying physics indicates that other spring mass system should also work, as long as the natural frequency is about 1 to 3 Hz, within the pace of a walk.

Another criterion for successfully doing the experiment/demonstration is related to visibility and contrast. Firstly, there should be large enough resonance amplitude. For instance, a system with a characteristic amplitude vs driving frequency curve as shown by the bottom dashed black curve in Fig. 3 is not a good choice. Secondly, the walking pace range should cover some out of resonance frequencies of the system. Or in technical words, the linewidth of the characteristic amplitude vs driving frequency spectrum of the system should be narrower than the frequency range of a typical walk. As such, a system with a characteristic amplitude vs driving frequency curve as shown by the upper dashed red curve in Fig. 3 is not an optimal choice. In short, this criterion guarantees a high contrast, a compelling difference between resonance oscillation and non-resonance oscillation.

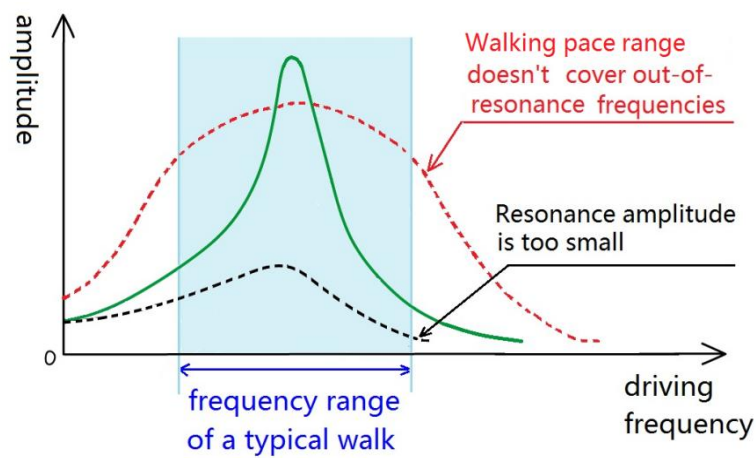

**Fig. 3. Illustration of the criteria for the proposed experiment.**

We further tried the experiment with a weight on a spring. With the weight being 0.10 kg and the spring constant being 10 N/m, the natural frequency of the system (and therefore also the predicted resonance frequency) is calculated to be 1.6 Hz. By analyzing the video, an obvious resonance phenomenon is observed while walking at a pace of 1.5 steps per second, which is consistent with the theory. This result support the criteria we have listed above and provide evidence to support our hypothesis concerning the underlying physics of the two spring scale oscillating system.

## 5. Discussion

Any system that meets the above criteria can be used for the demonstration. The two spring scale system and the weight on a spring system are two handy examples which can be set up using materials and equipment readily available in any school or university. The two spring scale system works better than the weight on a spring system in terms of visibility in a classroom.

It should be noted that in this paper, the term walking pace does not mean the same as walking speed. Instead, it refers to the number of steps per unit time, or frequency. It is the periodic motion of the feet that leads to a periodic motion in the vertical direction of the walker's body, which serves as the driving force on the spring scales. Therefore the forced oscillation of the spring scales is directly related to the frequency of the motion of the feet, but not the horizontal motion of the body.

The two spring scale system proposed here can be used as a short demonstration of forced oscillation and resonance and even for explaining the idea of frequency sweep. Before the walk, the teacher is advised to tell the students what he/she is going to do and then ask them to predict the behavior of the system. This will enhance the learning of the students [10-13]. It can also be used as a quantitative student experiment which involves 1) predicting and observing the behavior of the two spring scale system with different walking paces, and 2) finding the resonance frequency of a simple system and then comparing the experimental result with the theory.

## Acknowledgements

The author would like to thank the anonymous reviewers for their thoughtful and helpful suggestions which helped to improve the quality of this article significantly.